\documentclass[twocolumn]{revtex4}
\usepackage{amssymb}
\usepackage{latexsym}
\usepackage{epsfig}
\usepackage{amsmath}

\begin{document}
\title{Revisiting Holographic Dark Energy in Cyclic Cosmology}
\author{A. Sheykhi$^{1,2}$\footnote{asheykhi@shirazu.ac.ir},
 M. Tavayef$^{1}$, H. Moradpour$^{2}$\footnote{h.moradpour@riaam.ac.ir}}
\address{$^1$Physics Department and Biruni Observatory, College of
Sciences, Shiraz University, Shiraz 71454, Iran\\
         $^2$Research Institute for Astronomy and Astrophysics of Maragha
         (RIAAM), P.O. Box 55134-441, Maragha, Iran}

\begin{abstract}
Considering the holographic dark energy (HDE) with two different
Infrared (IR) cutoffs, we study the evolution of a cyclic universe
which avoids the Big-Rip singularity. Our results show that, even
in the absence of a mutual interaction between the cosmos sectors,
the HDE model with the Hubble radius as IR cutoff can mimics a
cosmological constant in the framework of a cyclic cosmology. In
addition, we find that both the interacting and non-interacting
universes may enter into a cycle of sequential contraction and
expansion, if the Granda-Oliveros (GO) cutoff is chosen as the IR
cutoff in the energy density of the HDE.
\end{abstract}

\maketitle
\section{Introduction}
There are a lot of observations data which signal us to an
accelerating expanding universe
\cite{perlmutter1999,riessAst1999,E.Kamatsu}. In order to describe
this current phase of the universe expansion, we need a strange
source of energy which is called dark energy and allocates
approximately $73\%$ of the cosmos tissue to itself. WMAP
observations also indicates that the pressure ($p$) and energy
density ($\rho$) of dark energy are bounded to
${p}/{\rho}\simeq-1.10$, a result addressing a phantom type dark
energy \cite{E.Kamatsu}. Therefore, in addition to the initial Big
Bang singularity, another singularity called Big Rip, can be
achieved by cosmos \cite{Caldwall.Kamionkowski,Barrow,S.Nojiri.
S.odintsov}. It was argued that a cyclic universe may avoid these
singularities \cite{Brown. Freese,Baum Frampton}.

In a cyclic cosmology, which has firstly been introduced by Tolman
\cite{Tolman}, our Universe successive expands and contracts. A
new version of cyclic model in which the universe cycle is
realized in the light of two separated branes, has been proposed
by Steinhardt et al. \cite{Turok}. Although, this new model gives
us an alternative to the inflation scenario, it suffers from two
problems including the entropy and black hole problems
\cite{Brown. Freese,Baum Frampton,X.Zhang}. It is finally useful
to note here that these problems can be solved by considering some
features of phantom dark energy \cite{Brown. Freese,Baum
Frampton}.

A cyclic universe is supported by an enormously high energy
density at the beginning and end of a cycle meaning that the
phantom dark energy increases more than expected in a finite time
\cite{Caldwall.Kamionkowski,caldwell2002}. As a result of this
behavior, one may expect that the high energy physics induces some
modifications to the Friedmann equation. Recently, such
modifications have been proposed arisen from braneworld scenario
\cite{Yshtanovsahni}, and Loop Quantum Cosmology (LQC) \cite{A.
Ashtekar2006,P. Singh2006}.

In LQC, which enumerates the quantum nature of spacetime \cite{A.
Ashtekar2006}, the Friedman equation is modified as
\begin{equation}\label{Friedeq01}
H^2=\frac{\rho}{3m_{ p}^2}\Big(1-\frac{\rho}{\rho_{c}}\Big),
\end{equation}
\noindent where $\rho$ is the total energy density of fluids which
support the background. Moreover, $H=\dot{a}/a$ and $m_{p}$ are
the Hubble parameter and the reduced planck mass ($m^2_{p}=1/8\pi
G=2.44\times10^{18}Gev$), respectively.
$\rho_{c}=4\sqrt{3}\lambda^{-3}m^{4}_{p}=0.82\rho_{p}$, where
$\rho_{p}=2.22\times10^{76}GeV$, denotes the critical energy
density set by quantum gravity and differs from the usual critical
density ($3 m_{p}^{2}H^{2}$). At the $\rho_c$ point, called
turnaround point, universe begins to collapse and by doing this,
it avoids the Big-Rip singularity. The universe contraction
continues until it reaches at the bounce point, where the universe
starts to expand again. It is worth mentioning that, at the
$\rho\ll\rho_{c}$ limit, the term $\rho/\rho_{c}$ in Eq.
(\ref{Friedeq01}) is negligible and thus the Friedmann equation is
obtainable.

There is a novel proposal for dark energy, which comes from
holographic hypothesis \cite{Bekenstein1994,L. Susskind}, and is
called holographic dark energy (HDE) \cite{cohen1999,Li2004}.
 This
proposal is based on the Cohen et al, idea where they suggest the
$L^{3}\Lambda^{4} \leq L _{m_{p}^{2}}$ relation between the
ultraviolet and IR cutoffs denoted as $\Lambda$ and $L$,
respectively \cite{cohen1999}.
 Applying this idea to cosmos, one
can get the holographic dark energy density as
\begin{equation}\label{HDEden}
\rho_{D}=3c^{2}m_{p}^{2}L^{-2},
\end{equation}
\noindent where $3c^{2}$ is a constant, and its value can be
determined by observations, and $L$ is a length which is related
to the size of the universe. A primary option for $L$ is the
Hubble parameter which does not lead to acceptable result in the
framework of standard cosmology  \cite{Hsu2008}. Assuming the
particle horizon as IR cutoff, the accelerated expansion cannot be
achieved, too \cite{Li2004}. To achieve an accelerated universe,
Li suggested that one should take the future event horizon of our
universe as the proper IR cutoff, in order to produce a correct
equation of state for HDE \cite{Li2004}. For a comprehensive list
of references concerning HDE, we refer to
\cite{zhang2006,nojiriOd2006,sheykhiphlet2009,sheykhi2010physlet,
sheykhi2009ph,setarej2010a,sheykhi2010,karamikhaled20011,Duranpav2011,sheykhi2012}
and references therein. The HDE model have also been constrained
and tested by various astronomical observations
\cite{shenwang2005,enqhan2005,zhang2009phy,michel2010}.

Furthermore, Granda and Oliveros have proposed another cutoff
which is a generalization of Ricci cutoff and is known as the GO
cutoff \cite{GO}. Some cosmological implications of the HDE model
with GO cutoff have been explored in
\cite{jamkasheykhi2012,Ghaffarisheydeh2014,Gaochen2009,wang}.

On the other hand, observations admit a mutual interaction between
the dark sectors of the universe
\cite{wang,gdo1,gdo2,ob1,ob2,ob3,ob4,ob5,ob6,ob7}. Motivated by
these observations, various kind of interactions have been
proposed and studied in various cosmological theories
\cite{Cai2010,Wei2011,WEI2011,chimen1,chimen2}. Finally, it is
worth mentioning that such interaction may solve the coincidence
problem \cite{pavonz,co1,co2,co3,co4,co5,co6}.

The studies on the HDE models with various cutoff were also
extended to a cyclic universe. In this regards, the cosmological
evolution of HDE model with the future event horizon radius as IR
cutoff have been investigated in a cyclic universe \cite{HDEcyc}.
Also, interacting new agegraphic dark energy in a cyclic universe
were explored in \cite{ADEcyc}. It was shown that when there is no
interaction between dark matter and dark energy, agegraphic dark
energy and new agegraphic dark energy do not produce a phantom
phase, and therefore cannot describe a cyclic universe
\cite{ADEcyc}. Furthermore, the ghost dark energy  and the
generalized ghost dark energy  in the framework of cyclic
cosmology were explored, recently \cite{sheykhitav}. It was argued
that in the absence of interaction in a cyclic universe, the
deceleration parameter becomes a constant; as a result the
universe cannot move from an accelerated expansion phase to a
deceleration phase and so cannot reach to\textit{ turnaround
point} and start to contract \cite{sheykhitav}. It was also argued
that only in the presence of interaction, the transition from an
accelerated expansion phase to a deceleration phase in the future
near the (\textit{ turnaround point}) can be achieved in a cyclic
universe \cite{sheykhitav}.

In the present paper, we are interested in studying the
cosmological consequences of the HDE hypothesis in the framework
of cyclic universe. It is noteworthy to mention that our work
differs from \cite{GO} in that, they studied HDE model with GO
cutoff in standard cosmology, while we investigate this model in a
cyclic universe. In order to achieve this goal, we focus on the
Hubble and GO cutoffs, and study the evolution of the universe.
The structure of this paper is as follows. In the next section, we
consider a cyclic universe filled with HDE and dark matter. As
system's IR cutoff we shall take the Hubble radius. In section
$\textmd{III}$, we study the HDE model in a cyclic universe with
GO cutoff. The last section is devoted to summary and concluding
remarks.
\section{HDE in a cyclic universe}
\subsection{Non-interacting case}
For a flat FRW universe filled by a pressureless matter and a dark
energy component, the Friedmann equation is
\begin{equation}\label{Fr}
H^2=\dfrac{\rho_{m}+\rho_{D}}{3m_{p}^{2}}\left[1-\left(
\dfrac{\rho_{m}+\rho_{D}}{\rho_{c}}\right)\right],
\end{equation}
\noindent where $\rho_m$ is the energy density of the pressureless
matter. In addition, when the cosmos sectors do not interact with
each other, the continuity equation leads to
\begin{equation}\label{coeq1}
\dot{\rho}_m+3H\rho_{m}=0,
\end{equation}
\begin{equation}\label{coeq2}
\dot{\rho}_{D}+3H \rho_{D}(1+\omega_{D})=0,
\end{equation}
\noindent where $\omega_{D}={p_D}/{\rho_D}$ and $p_D$ denote the
state parameter (EoS) and pressure of DE, respectively. Dot is
also the derivative with respect to cosmic time. We finally define
the dimensionless density parameters  as
$\Omega_{m}={\rho_{m}}/{(3m_{p}^{2}H^{2})}$ and
$\Omega_{D}={\rho_{D}}/{(3m_{p}^{2}H^{2})}$. Now, inserting
$u=\Omega_{m}/\Omega_{D}$ into Eq.~(\ref{Fr}), we obtain
\begin{equation}\label{rho/rhoc 2}
1-\frac{2\rho}{\rho_{c}}=\dfrac{2-\Omega_{D}\left(1+u \right)
}{\Omega_{D}\left(1+u \right)}.
\end{equation}
Considering the Hubble radius $L=H^{-1}$ as the IR cutoff, and
using Eq.~(\ref{HDEden}), one gets
\begin{equation}\label{denHDE/H-1}
\rho_{D}=3c^{2}m_{p}^{2}L^{-2}=3c^{2}m_{p}^{2}H^{2},
\end{equation}
and the density parameter becomes $\Omega_{D}=c^2$. Combining
Eqs.~(\ref{coeq2}) and (\ref{denHDE/H-1}), one arrives at
\begin{equation}\label{W/H-1}
\omega_{D}=-\frac{2}{3}\frac{\dot{H}}{H^{2}}-1.
\end{equation}
\noindent On the other hand, using
Eqs.~(\ref{Friedeq01}),~(\ref{coeq1}) and~(\ref{coeq2}), it is a
matter of calculation to show that
\begin{equation}\label{diffH/H2}
\frac{\dot{H}}{H^{2}}=-\frac{3}{2}\Omega_{D}\left(
1+u+\omega_{D}\right) \left(1-\dfrac{2\rho}{\rho_{c}} \right).
\end{equation}
\noindent Substituting this relation into Eq.~(\ref{rho/rhoc 2}),
 one gets
\begin{equation}\label{wH-1ni}
\omega_{D}
=\left[1-\Omega_{D}(1+u)\right]\left(\dfrac{1+u}{(1+u)(1+\Omega_{D})-2}\right).
\end{equation}
\noindent for the EoS of HDE in a non-interacting universe. It is
worth mentioning that, at the high energy limit, where
$\Omega_{m}\ll\Omega_{D}$ and thus $u\approx0$, this equation
leads to
\begin{equation}
\omega_{D}\approx-1.
\end{equation}
\noindent Indeed, for $u>0$, we always have $\omega_D<-1$, which
implies that a HDE with Hubble radius as IR cutoff, in a cyclic
universe, behaves as the phantom source. This result is clearly in
contrast to the non-interacting HDE model in standard cosmology
where $\omega_{D}=0$ for $L=H^{-1}$ \cite{Li2004}. Bearing in mind
the definition of deceleration parameter as
\begin{equation}\label{dq}
q=-1-\dfrac{\dot{H}}{H^{2}},
\end{equation}
\noindent and using Eqs.~(\ref{rho/rhoc 2}),~(\ref{diffH/H2}) and
(\ref{wH-1ni}) to evaluate the deceleration parameter, after
simple calculations, one finds
\begin{eqnarray}\label{qH-1}
q=-1+\frac{3u(2-\Omega_{D}(1+u))}{2[(1+u)(1+\Omega_{D})-2]}.
\end{eqnarray}
\noindent This implies that $q\approx-1$ at the $u\approx0$ limit.
This asymptotic behavior is an interesting result, because in the
original hypothesis of HDE, this IR cutoff leads to $q={1}/{2}$
\cite{Li2004}. Indeed, since $q\leq-1$ for $u\geq0$, the
accelerated phase of the universe expansion does not stop in this
model.
\subsection{Interacting case}
When the cosmos sectors interact with each other through a mutual
interaction $Q$, the energy-momentum conservation law is
decomposed as
\begin{equation}\label{coeqI1}
\dot{\rho}_m+3H\rho_{m}=Q,
\end{equation}
\begin{equation}\label{coneqI2}
\dot{\rho}_D+3H\rho_{D}(1+\omega_{D})=-Q.
\end{equation}
\noindent Throughout this paper, we consider an interaction term
as \cite{wang,pavonz}
\begin{equation}\label{QH-1}
Q=3b^{2}H\rho=3b^{2}H(\rho_{D}+\rho_{m})=3b^{2}H\rho_{D}(1+u),
\end{equation}
\noindent where $b$ is the coupling constant. Employing this
equation as well as Eqs.~(\ref{denHDE/H-1}) and~(\ref{coneqI2}),
one obtains
\begin{equation}\label{WI/H-1}
\omega_{D}=-1-\frac{2}{3}\frac{\dot{H}}{H^{2}}-b^{2}(1+u),
\end{equation}
\noindent and
\begin{equation}\label{diffH/H2I}
\frac{\dot{H}}{H^{2}}=-\frac{3}{2}\Omega_{D}\left(1+u+\omega_{D}\right)
\left(1-\dfrac{2\rho}{\rho_{c}}\right).
\end{equation}
\noindent wherein we combined Eqs.~(\ref{Fr}),~(\ref{coeqI1})
and~(\ref{coneqI2}) with Eq.~(\ref{WI/H-1}) to get the last
equation. One can also use Eqs.~(\ref{rho/rhoc 2}),~(\ref{WI/H-1})
and~(\ref{diffH/H2I}) in order to show that
\begin{eqnarray}\label{diffH/H2II}
\frac{\dot{H}}{H^{2}}=-\frac{3}{2}\Omega_{D}\left(
1+u+\omega_{D}\right)\left(\dfrac{2-\Omega_{D}\left(1+u \right)
}{\Omega_{D}\left(1+u \right) }\right),
\end{eqnarray}
\noindent and
\begin{equation}\label{wH-1}
\omega_{D}=\dfrac{(1+u)\left[1-(\Omega_{D}+b^{2})(1+u)\right]}{(1+u)(1+\Omega_{D})-2},
\end{equation}
\noindent At the $u\approx0$ limit, where $\Omega_{D}\approx
c^{2}$, we have
\begin{equation}\label{wD I H-1 2}
\omega_{D}\approx-1-\frac{b^{2}}{c^{2}-1},
\end{equation}
\noindent which is constant and smaller than $-1$. Therefore,
this model predicts a constant value for $\omega_D$ at the current
stage of the universe expansion, a result which is in conflict
with the recent observations which indicates that the DE candidate
should have a time varying EoS parameter \cite{alamsahni2004}.
Finally, calculations for the deceleration parameter lead to
\begin{equation}\label{qH-1Imain}
q=-1+\frac{3}{2}\left(2-\Omega_{D}(1+u)\right)\left[1+\frac{1-(\Omega_D+b^{2})(1+u)}{(1+u)(1+\Omega_{D})-2}\right],
\end{equation}
\noindent which finally yields
\begin{equation}\label{qH-1Iu=0}
q=2-\frac{3}{2}\Omega_{D}+\frac{3}{2}\frac{(2-\Omega_{D})(1-\Omega_{D}-b^{2})}{\Omega_{D}-1},
\end{equation}
\noindent for $u\approx0$, and thus the constant value of
\begin{equation}
q\approx\dfrac{(3b^{2}-2)c^{2}-6b^{2}+2}{2(c^{2}-1)},
\end{equation}
\noindent where we have used $\Omega_{D}= c^{2}$. If
$1<c^{2}<\dfrac{6b^{2}-1}{3b^{2}-2}$ and $b^{2}>\frac{2}{3}$, then
the deceleration parameter is negative, and therefore, our
Universe experiences an endless accelerating expanding phase
meaning that the \textit{turnaround point} is not achieved by our
Universe.
\section{HDE in a cyclic universe with GO cutoff}
Here, we consider the HDE model in a cyclic universe with GO
cutoff which has firstly been introduced by Granda and Oliveros as
\cite{GO}
\begin{equation}
L=(\alpha' H^{2}+\beta'\dot{H})^{-1/2},
\end{equation}
\noindent where $\alpha'$ and $\beta'$ are constants evaluated by
other parts of physics as well as observations. It is in fact the
formal generalization of the Ricci scalar curvature cutoff
\cite{Gaochen2009}, and a HDE model with this cutoff may avoid the
causality problem \cite{cau}. With this choice for cutoff, the
energy density (\ref{HDEden}) becomes
\begin{equation}\label{e}
\rho_{D}=3m_{p}^{2}(\alpha H^{2}+\beta \dot{H}),
\end{equation}
\noindent where $\alpha=c^{2}\alpha'$ and $\beta=\beta'c^2$, and
the dimensionless density parameter takes the form
\begin{equation}\label{omega}
\Omega_{D}=\frac{\rho_{D}}{3m_{p}^{2}H^{2}}=\alpha + \beta
\frac{\dot{H}}{H^{2}}.
\end{equation}
Finally, it is useful to note that since $c^2$, $\alpha$ and
$\beta$ are unknown coefficients, the GO cutoff can be redefined
as
\begin{equation}
L=(\alpha H^{2}+\beta\dot{H})^{-1/2},
\end{equation}
\noindent reduced to the Ricci cutoff ($L^{-2}=2H^{2}+\dot{H}$)
for $\alpha=2$ and $\beta=1$ \cite{Gaochen2009}.
\subsection{Non-interacting case}
For a non-interacting universe, substituting Eq.~(\ref{diffH/H2})
into Eq.~(\ref{omega}) and using Eq.~(\ref{rho/rhoc 2}), one
reaches at
\begin{eqnarray}\label{wHDEGO}
\omega_{D}=\left[
\frac{2}{3\beta}\left(\dfrac{\alpha-\Omega_{D}}{2-\Omega_{D}(1+u)}\right)-1\right]
(1+u).
\end{eqnarray}
\noindent In the high energy regime, where $u\approx0$, this
equation leads to
\begin{eqnarray}\label{wHDE2}
\omega_{D}\approx
\frac{2}{3\beta}\left(\dfrac{\alpha-\Omega_{D}}{2-\Omega_{D}}\right)-1,
\end{eqnarray}
\noindent which states that for $\alpha=2$ the Eos parameter, is
constant at this limit, namely $\omega_D=-1+2/3\beta$. In this
manner, in order to have a phantom source required for a cyclic
universe ($\omega_D<-1$), we should have $\beta<0$. Considering
the high energy limit where HDE dominates ($u\approx0$). In this
case we can neglect $\rho_m$ in the total energy density and the
Friedmann equation (\ref{Fr}) reduces  to
\begin{equation}\label{HDEC}
3m_{p}^{2}H^2=\rho_{D}\left(1-\frac{\rho_{D}}{\rho_{c}} \right).
\end{equation}
which can also be rewritten as
\begin{equation}\label{TT}
\Omega_{D}=1+\dfrac{\rho_{D}}{\rho_{c}-\rho_{D}},
\end{equation}
Inserting $\theta=\rho_{D}/\rho_{c}$ into Eq.~(\ref{TT}), one can
easily obtains
\begin{equation}\label{OMEGAtheta}
\Omega_{D}=1+\dfrac{\theta}{1-\theta}=\dfrac{1}{1-\theta},
\end{equation}
Combining with Eq.~(\ref{wHDE2}), we arrive at
\begin{eqnarray}
\omega_{D}=-1+\dfrac{2}{3\beta}\left(\dfrac{-\alpha \theta+
\alpha-1}{-2\theta+1}\right),
\end{eqnarray}
\noindent at the high energy limit. Therefore, for the Ricci
cutoff, where $\alpha=2$ and $\beta=1$, we have
$\omega_D={1}/{2}$.
\begin{figure}
\centering
\includegraphics[width=0.6\linewidth]{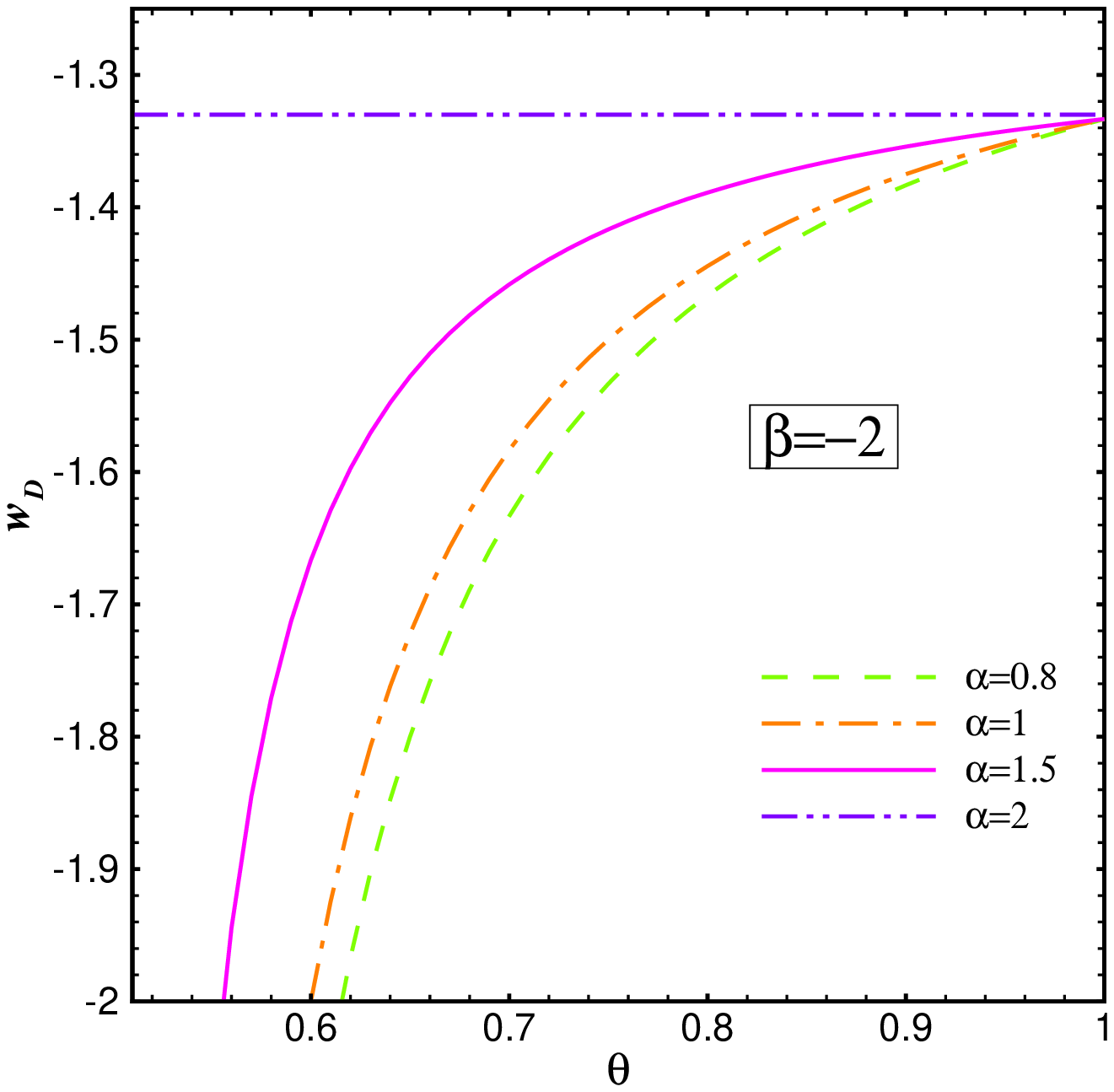}
\caption{The evolution of $\omega_D$ versus the dimensionless
quantity $\theta=\rho_{D}/\rho_{c}$, for different values of
$\alpha$ and $\beta=-2$.} \label{fig1} \centering
\includegraphics[width=0.6\linewidth]{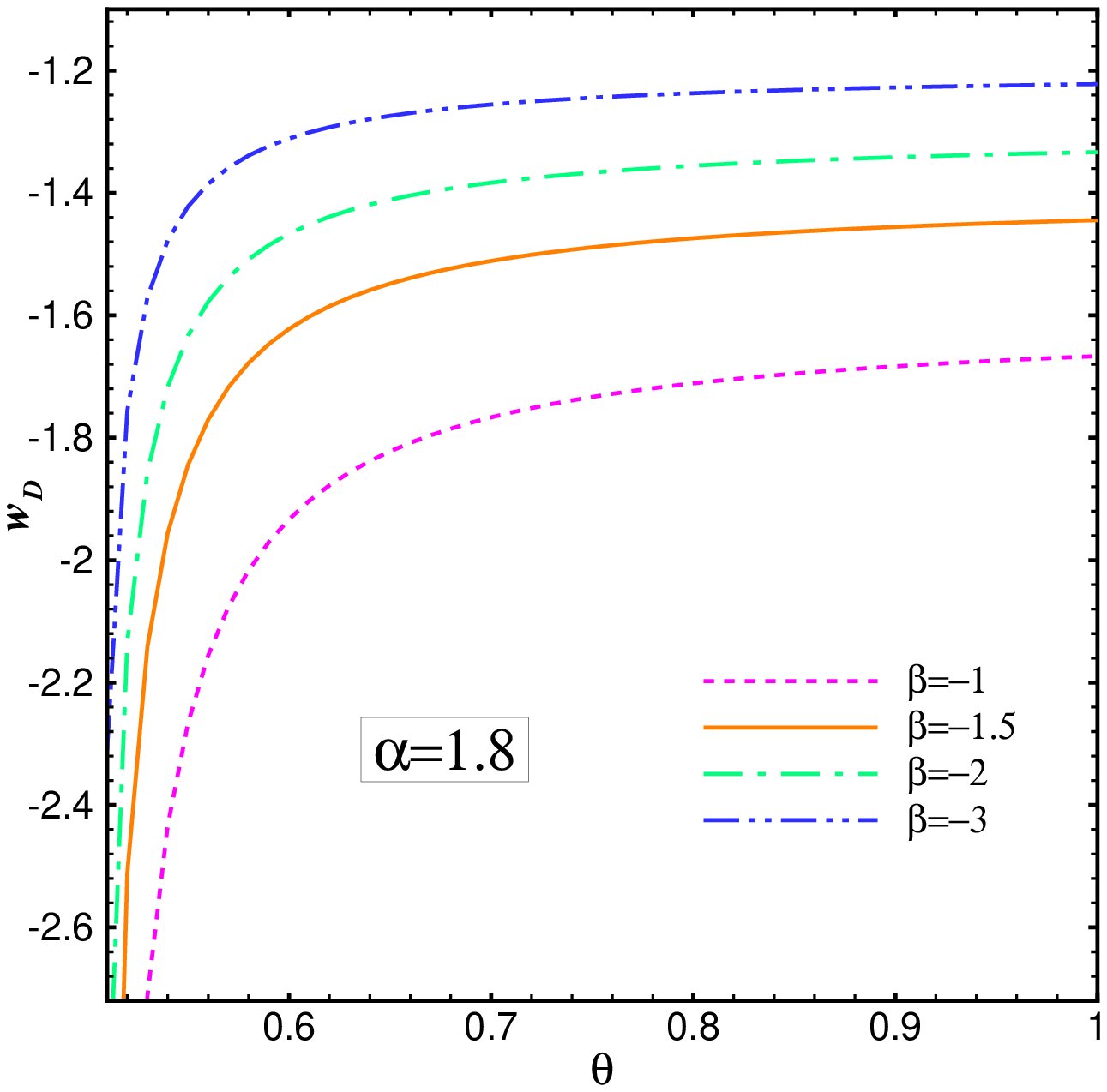}
\caption{The plot depicts $\omega_D$ as a function of $\theta$,
for different values of $\beta$, when $\alpha=1\cdot8$.}
\label{fig2}
\end{figure}
In the low energy regime in which $\rho\ll\rho_{c}$ and therefore
$3H^{2}=\rho$ meaning that the usual Friedmann equation is
obtainable, we have
\begin{equation}\label{stL}
1+u=\frac{1}{\Omega_{D}},
\end{equation}
\noindent where we have used Eq.~(\ref{rho/rhoc 2}) in order to
get this equation. Let us study the behavior of EoS
parameter~(\ref{wHDEGO}) at this limit, where the usual Friedmann
equation is available and $\Omega_{D}=1$ \cite{GO}. For this
propose, we substitute the above results into Eq.~(\ref{wHDEGO})
to obtain
\begin{equation}
\omega_{D}=-1+\frac{2(\alpha-1)}{3\beta}.
\end{equation}
\noindent This is  exactly the result obtained in \cite{GO} for
the EoS parameter of HDE with GO cutoff in standard cosmology.

In Fig.~\ref{fig1}, the evolution of the EoS
parameter, at high energy limit, is plotted as a function of
$\theta$ for $\beta=-2$ and  $\alpha=0\cdot8$, $1$, $1\cdot5$ and
$2$. As it is obvious, all diagrams converge to $ \omega_{D}\simeq
-1\cdot33 $ at turnaround point, and in fact, HDE behaves as the
phantom source for which $\omega_D<-1$. To complete the
discussion, the EoS parameter as a function of $\theta$ has also
been plotted for $\alpha=1\cdot8$ and different values of $\beta$
in Fig.~\ref{fig2} showing the effects of $\beta$ on
EoS parameter in high energy regime.

Now, using $\Omega_{D}^{\prime}={d\Omega_{D}}/{d
(\ln{a})}={\dot{\Omega}_D}/{H}$ as well as Eq.~(\ref{omega}), we
obtain
\begin{equation}\label{OmegaDHDE}
\Omega_{D}^{\prime}=\Omega_{D}\left(\frac{\dot{\rho_{D}}}{H\rho_{D}}-\frac{2\dot{H}}{H^{2}}\right).
\end{equation}
\noindent Additionally, one can combine
Eqs.~(\ref{coeq2}),~(\ref{diffH/H2}) and the above relation with
Eqs.~(\ref{wHDEGO}) and~(\ref{rho/rhoc 2}) to get the following
equation
\begin{equation}\label{OmegaDHDEMain}
\Omega_{D}^{\prime}=3\Omega_{D}\left[u+\frac{2(\alpha-\Omega_{D})}
{3\beta}\left(1-\dfrac{1+u}{2-\Omega_{D}(1+u)}\right)\right],
\end{equation}
\noindent which finally leads to
\begin{equation}
\Omega_{D}^{\prime}=\frac{2}{\beta}\Omega_{D}(1-\Omega_{D})
\left(\dfrac{\alpha-\Omega_{D}}{2-\Omega_{D}}\right),
\end{equation}
\noindent at the $u\approx0$ limit. Bearing Eqs.~(\ref{diffH/H2})
and (\ref{dq}) in mind, and using Eqs.~(\ref{rho/rhoc 2})
and~(\ref{wHDEGO}), we reach at
\begin{figure}
\centering
\includegraphics[width=0.6\linewidth]{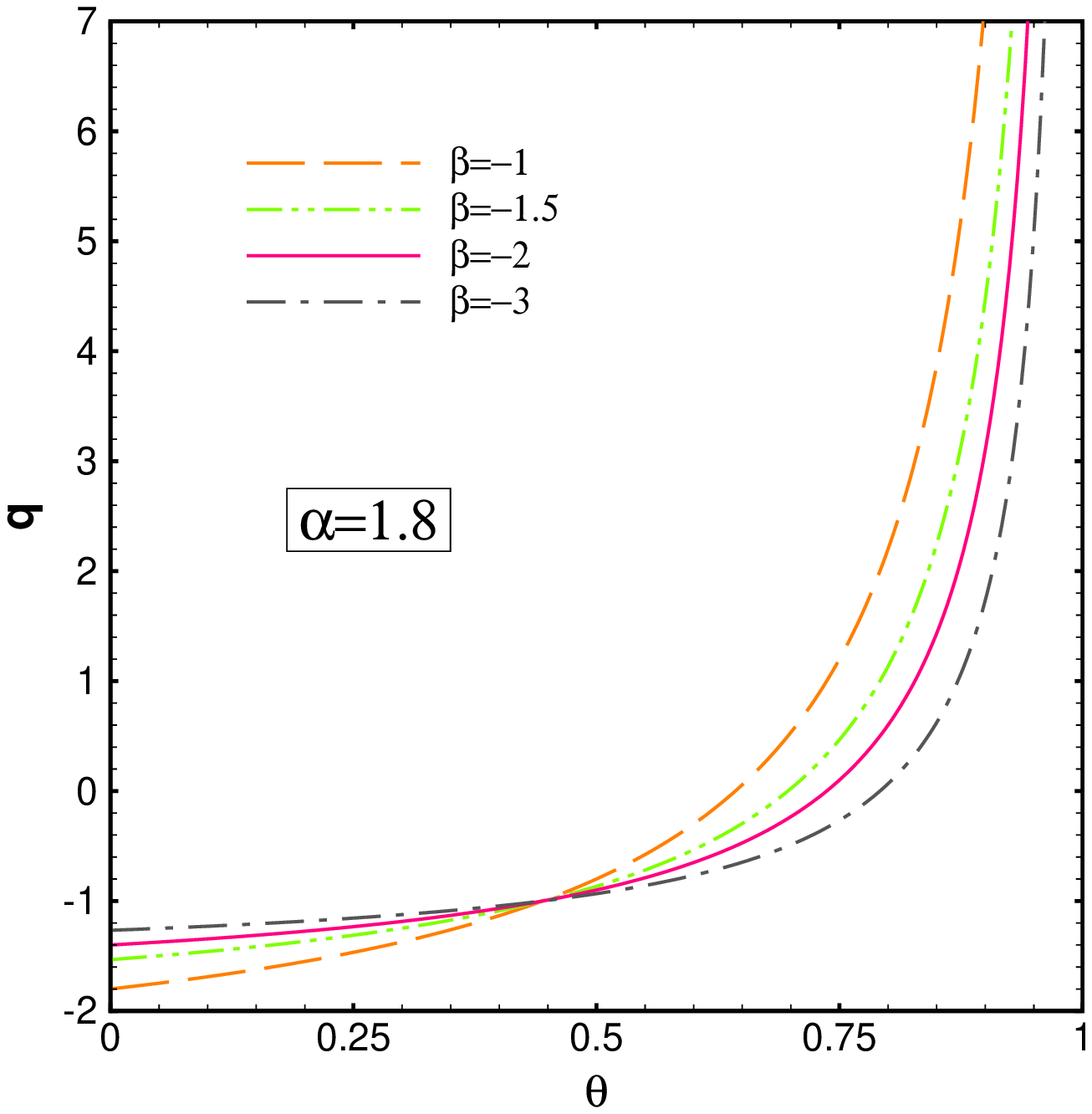}
\caption{The evolution of the deceleration parameter versus
$\theta$ for various values of $\beta$ whenever $\alpha=1\cdot8$.}
\label{fig3}
\includegraphics[width=0.6\linewidth]{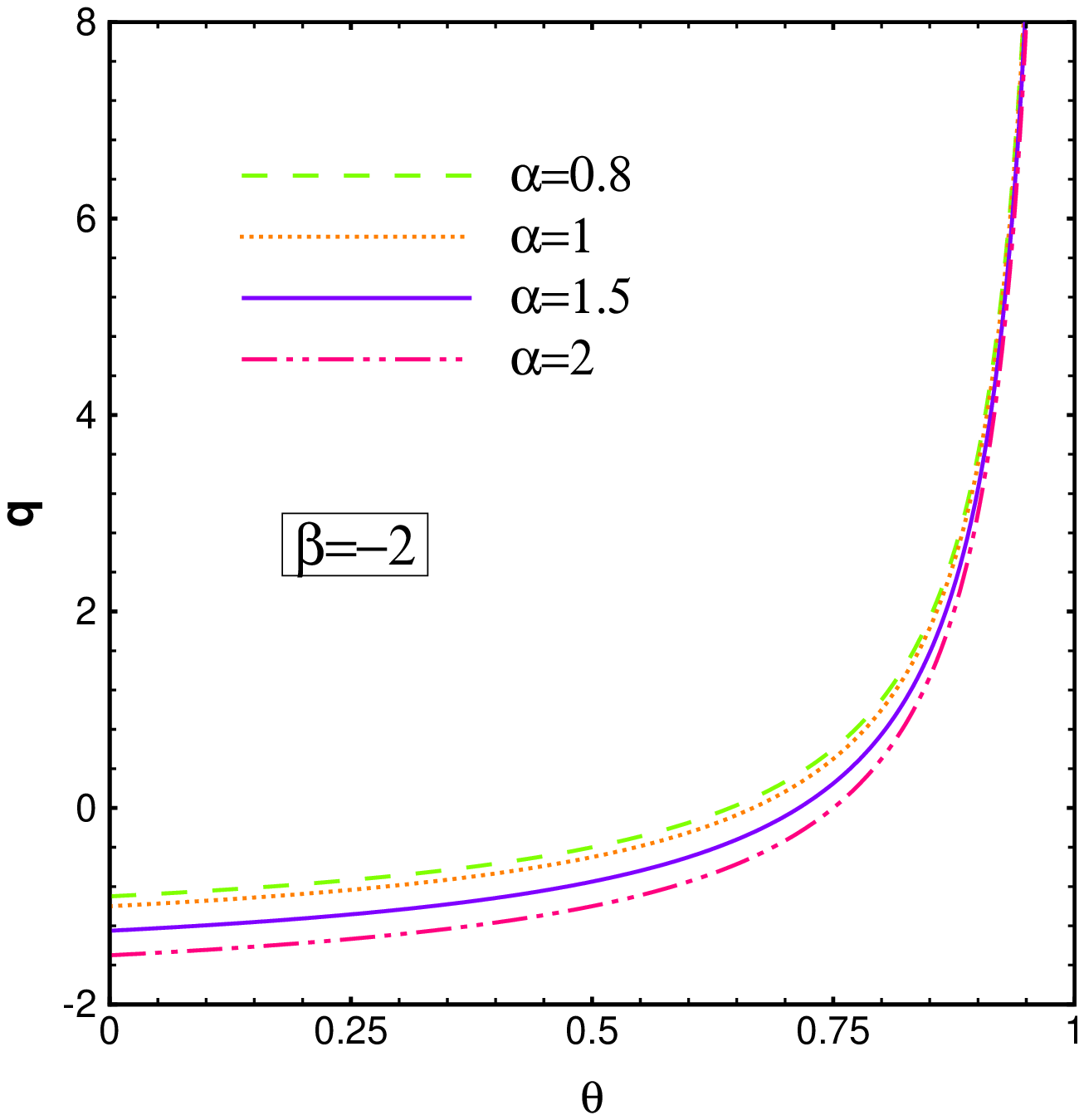}
\caption{The plot depicts $q$ as a function of $\theta$, when
$\beta=-2$, for some values of $\alpha$.}
\label{fig4}
\end{figure}
\begin{equation}
q=-1+\frac{\alpha-\Omega_{D}}{\beta},
\end{equation}
\noindent for the deceleration parameter. As we have already
discussed, in order to satisfy the phantom dark energy condition
i.e. $\omega_{D}<-1$, we should have $\beta<0$. In this case, an
accelerating universe ($q<0$) is achieved for
$\Omega_{D}<\alpha-\beta$. Besides, whenever
$\Omega_{D}=\alpha-\beta$, we have $q=0$ meaning that a transition
from an accelerated expansion phase to a decelerated one happens.
In Fig.~\ref{fig3}, the deceleration parameter as a
function of $\theta$ is plotted for various values of $\beta$ when
$\alpha=1.8$. The negative values of $q$ indicates that universe
experiences an accelerated expansion. Through the time and by
increasing the HDE phantom density, $q$ will become larger, until
it approaches to zero from below. Consequently, the universe moves
from an accelerated phase to a decelerating phase. It is worth
mentioning that this transition will earlier happen for smaller
values of $\beta$. Moreover, when the energy density approaches to
the critical energy density ($\theta\rightarrow1$), the universe
changes its evolution direction. We also plotted
Fig.~\ref{fig4} to investigate the effects of
$\alpha$ on $q$ showing that the larger values of $\alpha$ leads
to more negative values for $q$, and therefore, the foresaid
transition will happen later.
\subsection{Interacting case}
For an interacting universe, similar to the case of Hubble radius
cutoff in the previous section, taking the time derivative of
Eq.~(\ref{Friedeq01}), and combining the result with
Eqs.~(\ref{coeqI1}) and~(\ref{coneqI2}), we get
Eq.~(\ref{diffH/H2I}). Now, inserting this relation into
Eq.~(\ref{omega}) and using Eq.~(\ref{rho/rhoc 2}), one obtains
\begin{eqnarray}\label{wHDEGOI}
\omega_{D}=(1+u)\left[
\frac{2}{3\beta}\left(\dfrac{\alpha-\Omega_{D}}{2-\Omega_{D}(1+u)}\right)-1\right],
\end{eqnarray}
\noindent which is the same as Eq.~(\ref{wHDEGO}) in the
non-interacting case. Finally, for $u\approx0$ limit, the above
relation reduces to
\begin{eqnarray}\label{wHDE2I}
\omega_{D}=-1+
\frac{2}{3\beta}\left(\dfrac{\alpha-\Omega_{D}}{2-\Omega_{D}}\right).
\end{eqnarray}
On the other hand, combining Eqs.~(\ref{coneqI2})
and~(\ref{diffH/H2II}) with Eq.~(\ref{OmegaDHDE}), and inserting
the result into Eq.~(\ref{wHDEGOI}), we find
\begin{eqnarray}
\Omega_{D}^{\prime}&=&3\Omega_{D}\Bigg{\{}u-b^{2}(1+u)\\
\nonumber
&+&\frac{2(\alpha-\Omega_{D})}{3\beta}\left(1-\dfrac{1+u}{2-\Omega_{D}(1+u)}\right)\Bigg{\}}.
\end{eqnarray}
\noindent In the absence of interaction ($b^{2}=0$), our previous
result obtained in Eq.~(\ref{OmegaDHDEMain}) is recovered.
Moreover, in high energy regime where $u=0$, one gets
\begin{equation}\label{OmegaDHDEMain2}
\Omega_{D}^{\prime}=-3b^{2}\Omega_{D}+\frac{2}{\beta}\Omega_{D}(1-\Omega_{D})\left(\dfrac{\alpha-\Omega_{D}}{2-\Omega_{D}}\right).
\end{equation}
In order to achieve $q$, bearing Eq.~(\ref{dq}) in mind, and
substituting Eq.~(\ref{diffH/H2II}) into Eq.~(\ref{wHDEGOI}), one
can easily show that
\begin{equation}
q=-1+\frac{\alpha-\Omega_{D}}{\beta}.
\end{equation}
\noindent Therefore, unlike the evolution of the energy density
parameter Eq.~(\ref{OmegaDHDEMain2}), the deceleration parameter
of HDE model in a cyclic universe is independent of the coupling
constant $b^{2}$.
\section{Concluding remarks}
We have revisited the HDE model in the  framework of cyclic
cosmology by taking into account the Hubble radius as IR cutoff.
For a non-interacting universe, our study shows that, at the high
energy regime where $u=\rho_m/\rho_D\approx0$, both the EoS
parameter as well as the deceleration parameter of the HDE model
in a cyclic universe approach to $-1$. This implies that the
acceleration of the universe expansion can be achieved in this
model. Clearly, this result is in contrast to the HDE with Hubble
cutoff in standard cosmology where its EoS parameter behaves like
a dust, i.e., $\omega_D=0$, and hence cannot produce the
accelerating universe. We also considered the interacting HDE
model with Hubble radius as IR cutoff in the framework of cyclic
cosmology and found that in this case the EoS parameter crosses
the phantom divide, namely $\omega_D<-1$. It is worth noting that
in both of these model $\omega_D$ becomes a constant value at the
current stage of the universe expansion. This result is in
conflict with some recent observations which indicates that the DE
candidate should have a time varying EoS parameter
\cite{alamsahni2004}.

We also studied the HDE with GO cutoff in the framework of cyclic
universe. We found that this model supports a noninteracting
accelerating universe in the high energy limit. In this manner,
HDE behaves as a phantom source, and the universe expansion phase
will change from an accelerating phase to a deceleration one.
Finally, we found out that, for an interacting HDE with GO cutoff,
the mutual interaction between the dark sectors of cosmos only
affects the dimensionless density parameter of HDE, and it does
not affect the deceleration parameter meaning that the behavior of
$q$ in a noninteracting universe is the same as that of the
interacting universe.
\acknowledgments{We thank Shiraz University Research Council. This
work has been supported financially by Research Institute for
Astronomy \& Astrophysics of Maragha (RIAAM), Iran.}

\end{document}